# Relativistic plasma optics enabled by near-critical density nanostructured material

## Authors
J.H. Bin[1,2], W.J. Ma[1*], H.Y. Wang[1,2,3], M.J.V. Streeter[4], C. Kreuzer[1], D. Kiefer[1], M. Yeung[5], S. Cousens[5], P.S. Foster[5,6], B. Dromey[5], X.Q. Yan[3], J. Meyer-ter-Vehn[2], M. Zepf[5,7], & J. Schreiber[1,2†]

## Corresponding authors:
W.J. Ma & J. Schreiber

## Affiliations
[1]Fakultät für Physik, Ludwig-Maximilians-Universität München, D-85748 Garching, Germany
[2]Max-Planck-Institute für Quantenoptik, D-85748 Garching, Germany
[3]State Key Laboratory of Nuclear Physics and Technology, and Key Lab of High Energy Density Physics Simulation, CAPT, Peking University, Beijing 100871, China
[4]Blackett Laboratory, Imperial College London, SW7 2BZ, UK
[5]Department of Physics and Astronomy, Centre for Plasma Physics, Queens University Belfast, BT7 1NN, UK
[6]Central Laser Facility, STFC Rutherford Appleton Laboratory, Chilton, Didcot, Oxon, 0X11 0QX, UK
[7]Helmholtz-Institut-Jena, Fröbelstieg 3, 07743 Jena, Germany
*e-mail: wenjun.ma@physik.uni-muenchen.de
†e-mail: joerg.schreiber@mpq.mpg.de

**The nonlinear optical properties of a plasma due to the relativistic electron motion in an intense laser field are of fundamental importance for current research and the generation of brilliant laser-driven sources of particles and photons[1-15]. Yet, one of the most interesting regimes, where the frequency of the laser becomes resonant with the plasma, has remained experimentally hard to access. We overcome this limitation by utilizing ultrathin carbon nanotube foam[16] (CNF) targets allowing the strong relativistic nonlinearities at near- critical density (NCD) to be exploited for the first time. We report on the experimental realization of relativistic plasma optics to spatio-temporally compress the laser pulse within a few micrometers of propagation, while maintaining about half its energy. We also apply the enhanced laser pulses to substantially improve the properties of an ion bunch accelerated from a secondary target. Our results provide first insights into the rich physics of NCD plasmas and the opportunities waiting to be harvested for applications.**

The ratio of the plasma frequency $\omega_p$ to laser frequency $\omega_0$ separates different regimes of the interaction of lasers with plasma. Classically, $\omega_p/\omega_0 > 1$ and $\omega_p/\omega_0 < 1$ define overdense and underdense regimes respectively, in which the laser either can propagate or is reflected back from the plasma. Many intriguing phenomena are predicted or have been observed at the transition $\omega_p \approx \omega_0$ where the laser is resonant with the plasma. Studies have shown that the laser absorption increases, leading to strong electron heating[1] and generation of large magnetic fields[2-3]. The laser pulse is strongly modulated by relativistic self-focusing[4] and relativistic self-phase-modulation[5-6]. When a laser pulse irradiates a sufficiently thin overdense target with peak intensity beyond the relativistic threshold (~$10^{18}$ W/cm²), it can rapidly switch from opaque to transparent. This phenomenon of relativistically induced transparency[7] can lead to enhanced ion acceleration[8] or serve as ultrafast optical shutters[9]. NCD targets thus have great potential as a non-linear optical element by virtue of relativistic plasma effects to manipulate laser pulses to a desired shape at highest intensities[10]. This concept could be of great benefit for extensive investigations such as in plasma guiding devices[11], laser driven ion acceleration[12-13], single-cycle laser pulse generation[14], and coherent X- and Gamma-ray harmonic generation[15].

To exploit these phenomena in a controlled fashion, the challenge is to create sufficiently uniform, sharp-edged NCD plasma with lengths of several micrometers - the characteristic length scale on which the nonlinearities modify the pulse. In previous experiments, the NCD plasmas were provided either by expanding a solid-density target with an artificial[17] or inherent[12] prepulse of the laser or by using conventional foam material[18]. Both methods have severe limitations which we overcome by using ultrathin carbon nanotube foam[16] (CNF). CNF is an ideal candidate for creating thin NCD plasmas as detailed in Fig. 1. In particular its 3D network structure is homogeneous on scale of the laser wavelength (here $\lambda$=800nm). On the other hand, the intrinsic prepulse of a high intensity laser pulse, even with enhanced temporal contrast, will fully ionize the CNF and homogenize the plasma even further on a picosecond timescale prior to the main pulse. This results in a plasma slab with sharp boundaries and an average electron density $n_e$ ranged between $2n_c$ and $5n_c$, where $n_c = \varepsilon_0 m_e \omega_0^2 / e^2$ is the plasma critical density (for details see supplementary information).

To elucidate the concept of relativistic plasma optics at NCD, we conducted three-dimensional (3D) PIC simulations. The Astra-Gemini laser pulse (see Methods) interacts with a semi-infinite homogeneous plasma with critical density. Fig. 2a shows a 3D snapshot of the intensity distribution of the pulse at the time when it has self-focused from initially 3.5 µm full-width half-maximum (FWHM) diameter to about 1 µm. This happens over a propagation length of only 8 µm, in reasonable agreement with analytical estimates for relativistic self-focusing (see supplementary information). The maximum intensity increases by about a factor of 8 (Fig. 2b) and the leading edge of the pulse is steepened considerably. It rises from noise level to peak intensity over two laser cycles only (red curve in Fig. 2c). The steepening is caused by the temporal variation of refractive index in the NCD plasma and will be accompanied by spectral broadening and a positive chirp (for details see supplementary information). Integrating the laser intensity distribution over the temporal and spatial coordinates yields an energy transmittance of more than 50%.

**Experimental results**
The experiments were performed at the ASTRA Gemini laser at the Rutherford Appleton Laboratory in UK. The 50 fs Ti:Saphire laser pulse contains 4-5J laser energy after contrast enhancement with plasma mirrors, resulting in peak intensities of $2\times10^{20}$W/cm$^2$ which corresponds to the normalized vector potential amplitude $a_0 = 10$ (see Methods). The experimental set-up in Fig. 1d shows the diagnostics that were fielded simultaneously to characterize the interaction.

A second-order frequency-resolved optical gating (FROG) device allowed for temporal and spectral characterization of the transmitted laser pulses. The spectrum of the incident pulse, shown by the black curve in Fig. 3a, has a FWHM width of 23 nm corresponding to the nearly transform limited pulse duration of 50 fs (Fig. 3b). After propagation through a freestanding

CNF with an areal density of 3.5 μg/cm$^2$, the spectral width is significantly broadened to 37 nm (red curve in Fig. 3a). This broadening is required to support a steeper pulse. Further evidence can be inferred from the positive chirp which is observed in the Wigner distribution of the laser field in Fig. 3c. Indeed, the corresponding temporal profile shown by the overlaid blue curve is steepened. It is interesting to note that when using thicker CNF target, with areal density of 10 μg/cm$^2$, the temporal shape is almost unaltered (Fig. 3d) but overall spectrally blue-shifted (Fig. 3a). This indicates that in terms of pulse shaping the optimum thickness is around a few μg/cm$^2$. For thicker targets, the pulse steepening is less efficient, in agreement with previous theoretical work[10].

More details can be inferred from the measurements of laser energy transmittance through freestanding CNF. Fig. 3e compares this transmittance with results from nm-thick Diamond-Like-Carbon (DLC) foils[19] irradiated under the same conditions. In both cases the transmittance decreases with increasing areal density $\sigma = n_{e0}d_0$, where $n_{e0}$ and $d_0$ is the initial electron density and thickness of the material. The CNF-values remain significantly higher at any given areal density and reach up to 67%. Furthermore, simple analytical consideration (see Methods) shows that the transmittance of our CNF target is indeed consistent with propagation in the NCD regime.

**Application to laser driven ion acceleration.** Ion acceleration using intense lasers is an area of intense interest[20-24] and currently research aims to drive the development in two key areas: The maximum energy of ion beams and control of their spectral shape. Both benefit from increased laser intensity and utilization of very thin (nm) foils, which require high temporal contrast to accelerate ions efficiently. Consequently using CNF as active plasma optics to spatio-temporally compress the pulse is expected to enhance laser driven ion acceleration. To investigate their effectiveness, CNF layers with various thicknesses are directly deposited on nm-thin DLC foils (Fig. 4a).

Fig. 4b shows the proton spectra and C$^{6+}$ spectra for the linearly polarized Gemini laser pulses. The energy distributions monotonically decay and terminate at a maximum energy value. The maximum energy of protons increases from 12 to 29 MeV - a factor of 2.4 - with increasing thickness of the CNF layer (magenta dashed line), while the maximum energy per nucleon of carbon ions increases by a factor of ~1.7 (black dashed line). The observed increases are consistent with the significant intensity increase observed in the simulations (see supplementary information).

Using circularly polarized laser pulses results in remarkable changes in the ion spectra (Fig. 4c). Similarly to the linear case, the protons exhibit monotonically decaying spectral shapes. Now, the maximum energy of the protons increases by a factor of 1.5. The carbon-spectra behave notably different as they are elevated at energy around 70% of the maximum energy, similar to previous investigations[25]. More importantly, the energy per nucleon of the carbon ions increases much more rapidly than the proton energy with increasing CNF thickness (Fig. 4c), up to the point where they become comparable. Thus the velocities of both ion species are similar, a feature associated with a transition to ion acceleration dominated by the radiation pressure of the laser pulse (RPA)[25-29]. Since RPA is favored under conditions of high laser intensity and a sharp rising edge such a transition is consistent with the predicted pulse shaping in the CNF. It is worth noting that the maximum energy of the carbon ions is increased by a factor of 2.7 for the largest CNF thickness under investigation, corresponding to ~20 MeV/u. This is to the best of our knowledge, the highest value for carbon ions demonstrated from a Gemini-class laser system to date.

In conclusion, we have demonstrated the potential and practical feasibility of spatial and temporal modification of laser pulses at high intensities by exploiting relativistic optics in near-critical density targets. This principle is already crucial to many recent advances in low density plasmas, such as laser wakefield acceleration, where relativistic self-focusing and self-phase modulation has enabled revolutionary progress over the last decade[30]. Likewise, we succeeded to increase ion energies by a factor of about 3 in our 'proof-of-principle' experiment, which would otherwise require much larger lasers. The introduction of controlled NCD plasmas thus allows relativistic non-linearities to be exploited at highest intensities.

**Acknowledgements**
The work was supported by the DFG Cluster of Excellence Munich-Centre for Advanced Photonics (MAP) and the Transregio TR18, as well as the International Max Planck Research School of Advanced Photon Science (IMPRS-APS). The experiment was funded by EPSRC und the A-SAIL and LIBRA grants. We thank the staff of the ASTRA Gemini operations team and the CLF for their assistance during the experiment.

**Author Contributions**
J.H.B., W.J.M., J.S. and M.Z. prepared the initial manuscript. All authors contributed revision to the final version. H.Y.W. performed the PIC simulations. W.J.M. prepared the targets. Data analysis was carried out by J.H.B., M.J.V.S. and W.J.M. All authors contributed to the planning, implementation and execution of the experiments.

**Competing financial interests**
The authors declare no competing financial interests.

# Figures

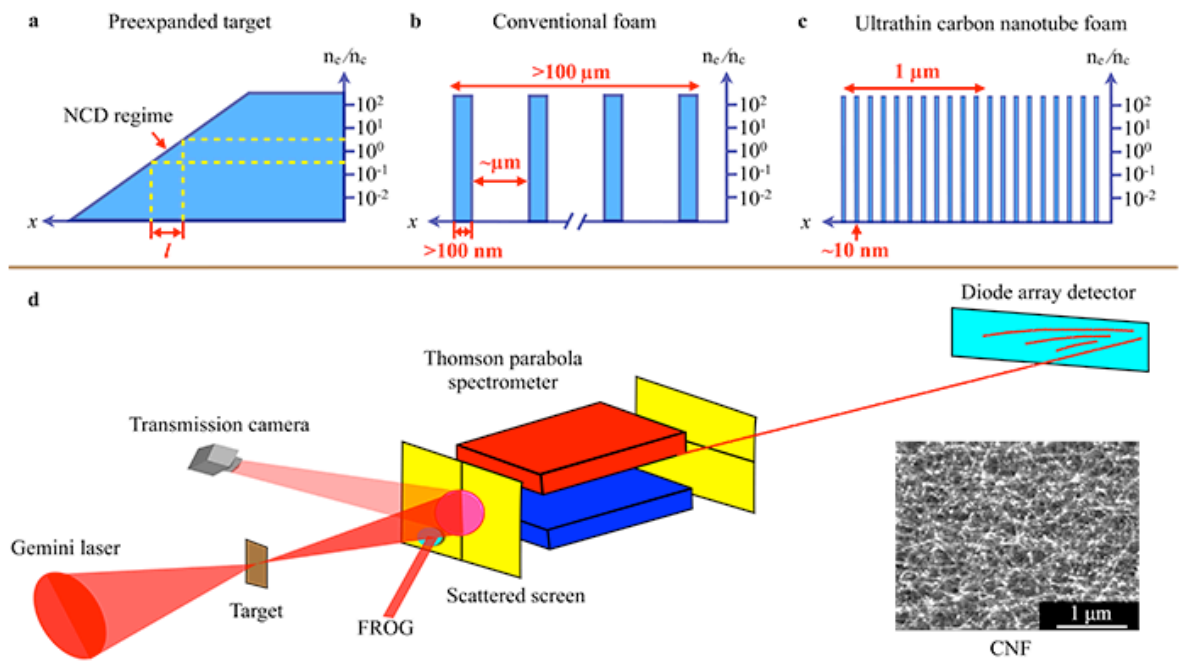

**Figure 1 | Schematic of NCD targets and experimental set-up. a**, NCD plasma with long scale-length and short NCD region typical for the production by pre-expansion of a solid density target. **b**, Conventional foam targets are composed of high-density material interspersed by vacuum with a non-uniformity on length scales comparable or greater than the laser wavelength. **c**, Ultrathin carbon nanotube foam (CNF) is deposited as a thin foam, or haystack composed of carbon nanotube bundles. Each nanotube bundle has a diameter of ~10 nm and adjacent carbon nanotube bundles are spaced by about 50~100 nm. The resulting web-like structure is homogeneous on scale of the laser wavelength. **d**, In the experiment, the Gemini laser pulse is normally incident on the target. A Thomson parabola (TP) spectrometer along with a diode array (DA) detector was used to detect the ions under target normal. The time-integrated transmitted laser profile is detected with a scattered screen placed in front of TP and a charge coupled device (CCD) camera. One frequency-resolved-optical-gating (FROG) device is implemented to diagnose the laser pulse transmitted through the target. The inset shows a scanning electron microscopy (SEM) image of CNF.

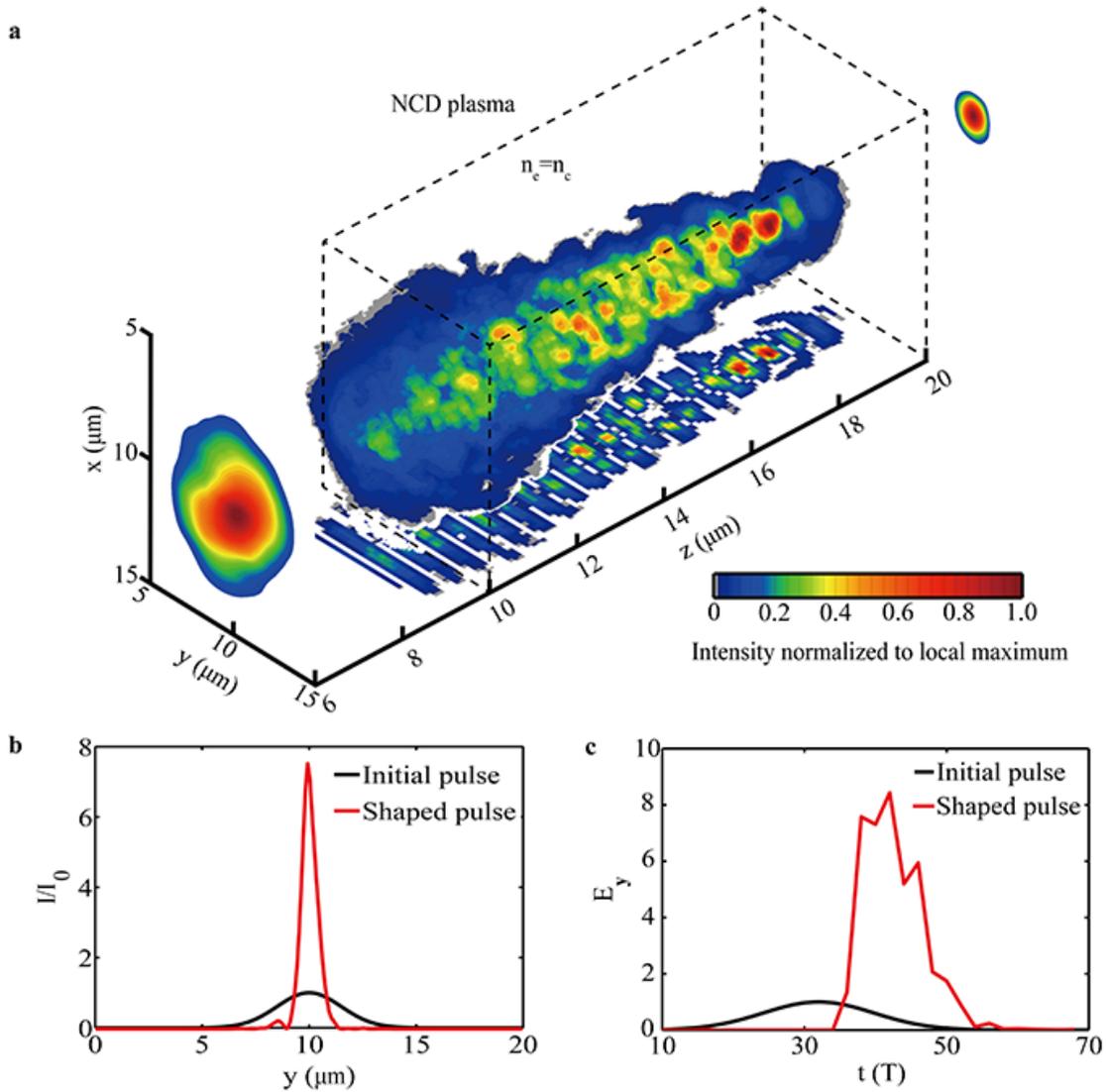

**Figure 2 | Three-dimensional particle-in-cell (PIC) simulation of Gemini-laser interaction with critical density plasma. a**, Perspective snapshot showing the laser pulse undergoing self-focusing in the NCD plasma. Also shown are the corresponding transverse cross-sections at the plasma entrance (black dashed at z=10 µm) and at z= 18 µm, respectively and the intensity distribution projected to the y-z plane. For better illustration the intensities are normalized to the local maximum along the propagation direction. **b**, Transverse lineout of the laser intensity at z= 18 µm (red curve) in comparison to the initial transverse laser profile at the plasma entrance at z=10 µm (black curve) showing an 8-fold increase of intensity. **c**, On axis longitudinal, i.e. temporal laser profile at z=18µm (red curve) and the initial laser pulse (black curve). For the details of simulation parameters see the methods.

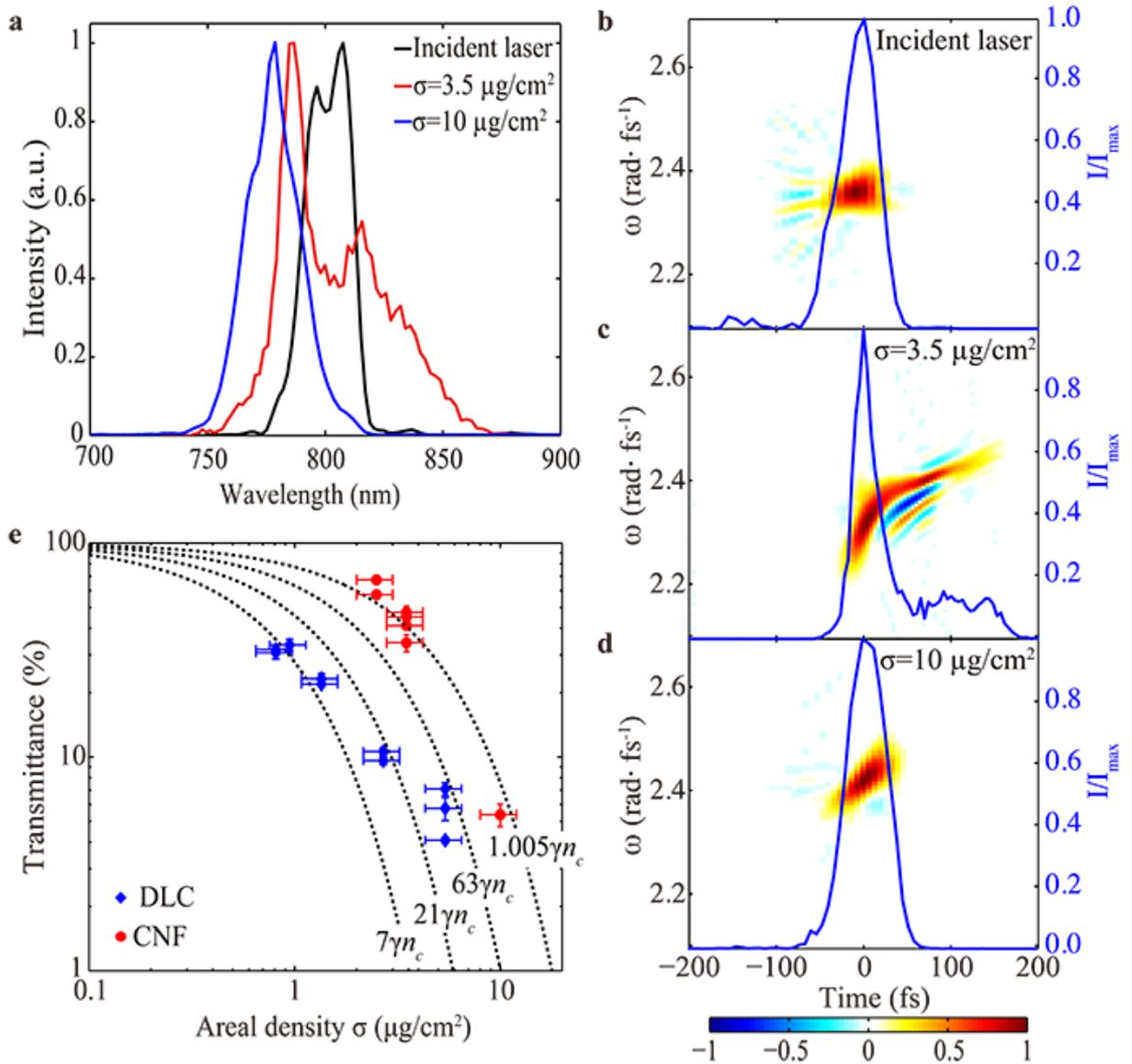

**Figure 3 | Experimental characterization of the laser pulses transmitted through the CNF targets. a**, Normalized laser spectra of incident laser (black line) compared to transmitted through CNF targets with areal density $\sigma = 3.5 \mu g/cm^2$ (red) and $\sigma = 10 \mu g/cm^2$ (blue). **b**, Wigner distribution of fields retrieved from the FROG-measurements for incident laser, **c**, for the CNF targets with $\sigma = 3.5 \mu g/cm^2$, and **d**, with $\sigma = 10 \mu g/cm^2$. The overlaid blue curves in **b**, **c**, and **d** represent the temporal intensity distribution ($t<0$ is early). **e**, Time-integrated laser transmittance for the Gemini laser pulse, interacting with DLC (blue squares) and CNF (red circles) targets, are plotted as a function of the areal density $\sigma$. The black dashed curves correspond to the transmission of a collisionless plasma with electron density $n_e = 63\gamma n_c$, $21\gamma n_c$, $7\gamma n_c$, and $1.005\gamma n_c$, respectively. Here $\gamma = \sqrt{1+(a_0/2)^{1/2}} \approx 7$. Note that while the transmittance of the CNF closely follows a theoretical curve for a single density ($1.005\gamma n_c$), the transmittance of the DLC implies that thinner foils have a lower density at the time of peak intensity. This is due to decompression induced by the rising edge of the pulse (see supplementary information).

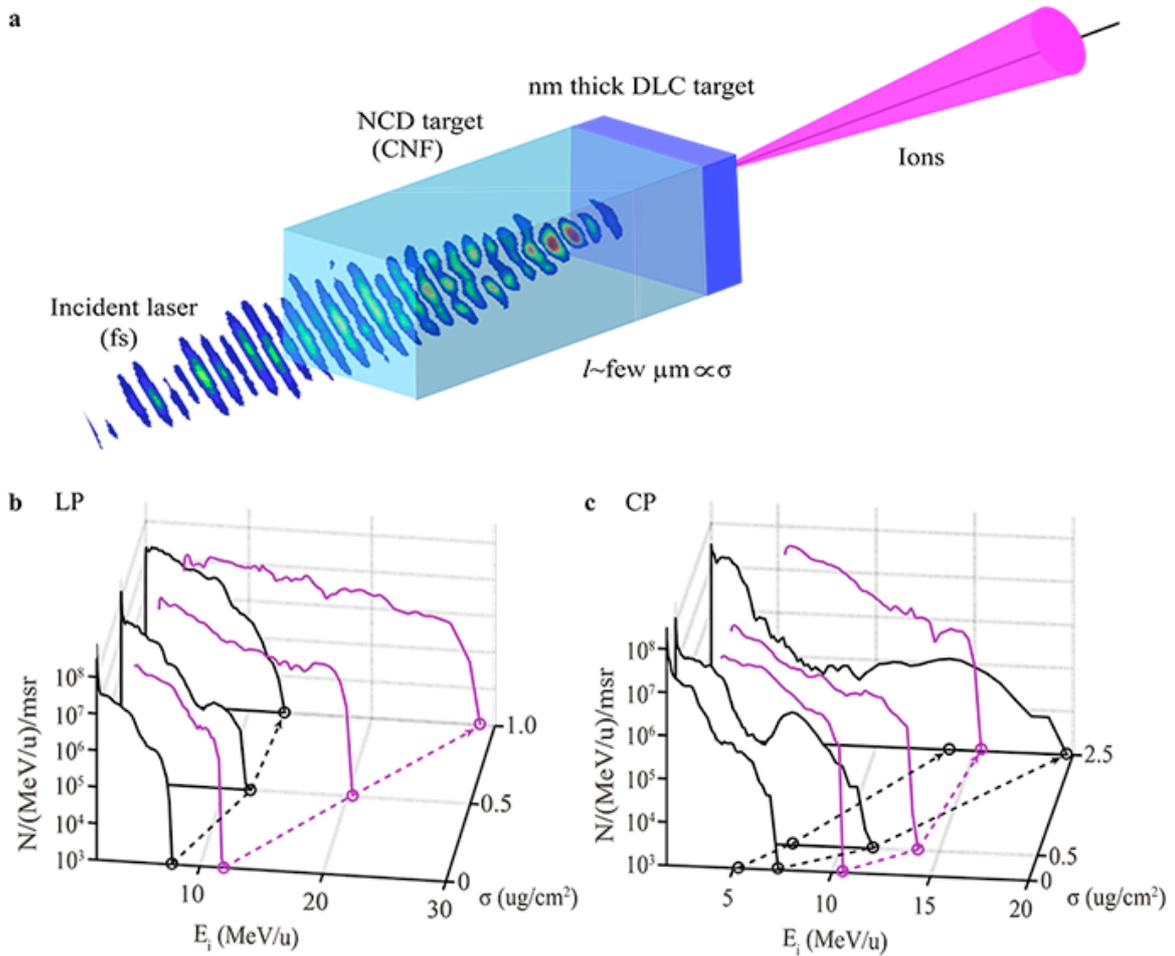

**Figure 4 | Application of relativistic plasma optics to laser driven ion acceleration. a**, The CNF target (i.e. the plasma optics) is directly attached to the second DLC foil acting as the actual ion source. **b**, Ion energy distributions for the case of linearly polarized laser pulses interacting with CNF attached to a 20 nm DLC foil, and **c**, for circularly polarized laser pulses interacting with CNF attached to a 10 nm DLC foil. The magenta curves show the proton spectra, the black curves the corresponding $C^{6+}$-ion spectra from the same laser shot. The dashed lines in **b** and **c** visualize the dependence of representative energy parameters, in particular the maximum energy and the energy of the spectral peak on the length of the CNF, represented by the areal density σ in µg/cm². The trends to increased energies with increasing CNF-thickness are evident, for carbon ions a factor of 2.7 is observed.

# Methods

**Laser system.** The experiments were performed at the ASTRA Gemini laser at the Central laser facility of the Rutherford Appleton Laboratory in the UK. The system delivers pulses with duration of 50 fs FWHM and central wavelength at 800 nm. A re-collimating double plasma mirror system was introduced to further enhance the laser contrast with a typical energy loss of 50 percent. A f/2 off-axis parabolic mirror (OAP) is used to focus the pulses to a measured FWHM diameter of 2.7×4.9 µm. Due to losses in the beamline and the plasma mirrors as well as imperfections in the focal spot, we achieve a peak laser intensity of $2\times10^{20}$ W/cm$^2$ ($a_0$~10) with 4-5 Joules of laser energy.

**Particle-in-cell simulations.** Three-dimensional simulations were carried out with the KLAP3D particle-in-cell code (ref. 10, ref. 27). The 3D simulation box size is $20\mu m \times 20\mu m \times 40\mu m$ with a resolution of 20 cells/µm. A circularly polarized laser pulse with a Gaussian temporal profile with an intensity-envelope FWHM duration of 50 fs and a transverse Gaussian profile with a FWHM diameter of 3.5 µm modeled the ASTRA Gemini laser pulse. We set $a_0 = 7$ for circular polarization, corresponding to a peak laser intensity of $I_0 = 2 \times 10^{20} W/cm^2$. Note that in the simulation, the minimum intensity in the initial pulse is of $10^{-4}$ of the peak intensity. A uniform carbon plasma layer of density $n_e = n_c$ is placed between 10 µm to 40 µm with initial electron temperature of 1 keV.

**Transmittance calculations.** For a collision-less, box-shaped plasma with thickness $d$ and density $n_e$, the transmittance is given by $T = \exp(-2d/l_s)$, where $l_s = \lambda_0 \cdot \left(2\pi\sqrt{\frac{n_e}{\gamma n_c}-1}\right)^{-1}$ is the skin-length. The skin-length depends on laser wavelength $\lambda_0$ and the corresponding critical density $n_c$, which is corrected for the relativistic motion of the electrons in the strong laser field. For $a_0 = 10$ we estimate the cycle-averaged $\gamma = \sqrt{1+a_0^2/2} \approx 7$.

We introduce the areal density $\sigma = n_{e0}d_0$ defined as the product of the initial electron density of the plasma $n_{e0}$ and its initial thickness $d_0$. Assuming that the shape of the plasma remains box-like, the areal density remains constant, even when premature expansion occurs, i.e. $\sigma = n_{e0}d_0 = n_e d$. Such pre-expansion is well-known in the case of DLC-foils. The transmittance can then be written as

$$T = \exp\left\{-4\pi \frac{\sigma_0}{n_e \lambda_0} \sqrt{\frac{n_e}{\gamma n_c}-1}\right\}.$$

Starting from a highly overdense plasma ($n_e \gg 2\gamma n_c$), the transmittance decreases with decreasing electron density until it reaches the minimum for $n_e = 2\gamma n_c$. Thereafter, it rapidly rises and reaches 1 for $n_e = \gamma n_c$. The apparent electron density at the time of interaction with the peak intensity depends on the initial thickness in the case of DLC foils (see supplementary Information), therefore, the acquired data points cross transmittance curves calculated for different apparent densities $n_e$. In contrast, the CNF data are consistent with a single density value which is just slightly (relativistically) over-dense.

# Supplementary information for "Relativistic plasma optics enabled by near-critical density nanostructured material"

**Authors**
J.H. Bin[1, 2], W.J. Ma[1], H.Y. Wang[1, 2, 3], M.J.V. Streeter[4], C. Kreuzer[1], D. Kiefer[1], M. Yeung[5], S. Cousens[5], P.S. Foster[5, 6], B. Dromey[5], X.Q. Yan[3], J. Meyer-ter-Vehn[2], M. Zepf[5, 7], & J. Schreiber[1, 2]

**Affiliations**
[1]Fakultät für Physik, Ludwig-Maximilians-Universität München, D-85748 Garching, Germany
[2]Max-Planck-Institute für Quantenoptik, D-85748 Garching, Germany
[3]State Key Laboratory of Nuclear Physics and Technology, and Key Lab of High Energy Density Physics Simulation, CAPT, Peking University, Beijing 100871, China
[4]Blackett Laboratory, Imperial College London, SW7 2BZ, UK
[5]Department of Physics and Astronomy, Centre for Plasma Physics, Queens University Belfast, BT7 1NN, UK
[6]Central Laser Facility, STFC Rutherford Appleton Laboratory, Chilton, Didcot, Oxon, 0X11 0QX, UK
[7]Helmholtz-Institut-Jena, Fröbelstieg 3, 07743 Jena, Germany

**Ultrathin carbon nanotube foams**

Ultrathin carbon nanotube foams (CNF) are fabricated through floating catalyst vapor deposition (FCCVD)[1]. Ferrocene/sulfur powder is sublimated under 85 °C as catalyst, and carried into the reaction zone by a gas mixture of 1400 sccm argon and 10 sccm methane. The temperature of the reaction zone is 1100°C. In the reaction zone, single walled carbon nanotubes grow out of the floating catalyst to a length of tens of micrometers within a few seconds, and self-assemble as bundles with a diameter of 10~20 nm. After their growth, nanotube bundles are transported to the deposition zone by the carrier gas, and deposited as a thin foam. The spaces between adjacent carbon nanotube bundles is about 50~100 nm. Consequently, the foams are highly homogeneous above micrometer scale. The thickness of the foam is controlled by the deposition time, and its average density is controlled by the feeding rate of catalyst and methane. Under the parameters given above, the growth rate of the foam is 0.2 μg/cm$^2$/min. The average density of the foam is 20±10 mg/cm$^3$, which corresponds to a plasma density of 3.5±1.5 $n_c$ if all the carbon atoms are fully ionized.

CNFs are a highly absorptive material for lasers with center wavelength of 800 nm. Thus it has a much lower damage threshold as compared to transparent materials, for example, Diamond-like carbon (DLC) foils. Supplementary Fig. 1 shows the experimental measurement results of damage threshold for CNF with varying laser pulse duration. The prepulse of the Gemini laser reaches the intensity of 10$^{10}$ W/cm$^2$ at around 5 ps, where CNFs are then ionized and start to expand. With a typical ion sound velocity $c_s$~2.2 × 10$^6$ cm/s, a single carbon nanotube bundle would evolve into a plasma with scale length of about 100 nm within 5 ps[2]. This expansion results in a merging of 3D CNF structure and therefore improves the homogeneity of the near-critical density slab even further.

**Analytical estimate of the self-focused spot size and length**

The underlying physics of relativistic self-focusing can be understood in terms of the spatial variation of refractive index due to the relativistic motion of the electrons in the strong laser field. The refractive index of a plasma can be expressed as $\eta = ck/\omega = (1 - n_e/\gamma n_c)^{1/2}$. The term $\gamma = (1+a^2)^{1/2}$ represents the relativistic mass correction due to the motion of the plasma electrons in a circularly polarized laser field with normalized vector potential $a$. Consider a laser beam with Gaussian radial intensity profile $a(r) = a_0 \exp(-2\ln 2 \cdot r^2/D_L^2)$ with wavelength $\lambda_L$ interacting with uniform plasma. $D_L$ is the full-width half maximum (FWHM) diameter of the laser. Thus the refractive index is larger in regions of high laser intensity. Hence the plasma acts as a focusing lens and focuses the laser to a smaller spot size, i.e., relativistic self-focusing occurs. The phase velocity can be approximated as $v_p(r) = c/\eta(r) \approx c \cdot (1 + n_e/2n_c a(r))$. The approximation is valid for large laser intensities ($a(r) \gg 1$). The maximum difference of velocity through the area of $\sqrt{2}D_L$ (FWHM diameter of $a(r)$) is $\Delta v_p = cn_e/2n_c a_0$. Thus the divergence angle of the laser beam is given by:

$$\theta = (\Delta v_p/c)^{1/2} = (n_e/2n_c a_0)^{1/2}$$

With this, the spot FWHM diameter $D_{FWHM}$ after propagation over the self-focusing length $f$ can be estimated by applying the Gaussian beam propagation as

$$D_{FWHM} = 2\sqrt{\ln 2} \cdot \lambda_L/\pi\theta = 2\lambda_L \cdot (2\ln 2 n_c a_0/n_e)^{1/2}/\pi \approx 0.74 \lambda_L \cdot (n_c a_0/n_e)^{1/2} \quad (1)$$

$$f = \sqrt{2}D_L/2\theta = D_L \cdot \left(\frac{n_c a_0}{n_e}\right)^{1/2} \approx D_L \cdot (n_c a_0/n_e)^{1/2} \quad (2)$$

Except for some small deviation in pre-factors, this is the same scaling for the focal spot size as obtained in previous works[3, 4].

For the 3D simulation parameters in the main manuscript, Eq. (1) and (2) predict that the initial laser will self-focus to a focal spot with a FWHM diameter of 1.5 µm after a propagation distance of 9.3 µm in a critical density plasma. This agrees fairly well with the values extracted from the simulation. Moreover, Eq. (1) also reveals that the strongest self-focusing occurs in the NCD regime since $D_{FWHM}$ is inversely proportional to the square root of the initial electron density. Of course, a density considerably larger than the (relativistic) critical density $\gamma n_c$ is not of interest as the laser would not propagate.

**Steepening of the pulse front**

The steepening of the pulse front is caused by the time dependency of the refractive index $\eta$. At the beginning, the target is opaque and the laser pulse can only penetrate evanescently into the target characterized by the skin depth $l_s$ because $\eta$ is imaginary (the initial electron density $n_e > n_c$). As the laser intensity and the $\gamma$-factor of the electron increase during the interaction, the plasma eventually becomes transparent when $n_e/\gamma \leq n_c$. Thus, $\eta$ becomes real and the target suddenly acts as a transparent medium. This transition from opaque to transparent is the first contribution to the steepening of the rising edge of the pulse. Thereafter whilst propagating the intensity dependent refractive index causes relativistic self-phase-modulation. In simple words, the group velocity $v_g$ at the pulse front is lower such that the center of the pulse can catch up with it which further steepens the pulse as observed in the simulation (see the main manuscript Fig. 1e). The signature of pulse steepening in the spectral domain is spectral broadening, but also redshift at the front and blueshift at the back of the transmitted pulse, i.e., a positive chirp. Those signatures are observed in our simulation as well as in the experiments.

**Pre-expansion induced reduction of electron density of DLC foils**

The observed transmission values for DLC foils with various initial thicknesses cannot be explained by a single density value $n_e$. This is due to the fact that they are ionized long before the laser pulse reaches its peak intensity and then expands for certain time. We assume that the expansion velocity is independent of the initial thickness $d_0$ of the plasma[5]. For the sake of simplicity we assume that the density distribution along the target normal remains essentially box-like. Then the electron density at the time of peak intensity interaction, which we extract from Fig. 3e, is given by $n_e = \dfrac{n_0 d_0}{d_0 + 2d_{\exp}}$, where $n_0 = 450 n_c$ for completely ionized DLC and $d_{\exp}$ is a fixed expanded thickness value for a given degree of expansion. Supplementary Fig. 2 shows good consistence with this very simplistic model.

**Intensity increase inferred from increased ion energies**

The ions are accelerated in an electric field set up by fast laser-accelerated electrons. The actual dynamics, i.e. dominated by plasma expansion or alternative schemes currently under investigation, plays a minor role for our discussion. In the experiment we observe monotonically decreasing spectra which terminate at a maximum energy. With increasing CNF layer thickness this maximum energy increases from 12 to 29 MeV - a factor of 2.4. At the same time, the maximum energy per nucleon of the $C^{6+}$-ions increases by a factor of 1.7 ($\sim 2.4/\sqrt{2}$). This suggests that the acceleration potential ΔV is increased by a factor of 2.4 for both ion species. Depending on the parameter range, in target normal sheath acceleration $\Delta V$ scales with the square-root[6,7,8] of or even linear[9] with the laser intensity $I_0$. Thus, the CNF foam increases the laser intensity by a factor of 2.4…5.8. In the break-out-after burner regime, $\Delta V \propto I_0^{1/2} t_{FWHM}^{1/3}$ also depends weakly on the laser pulse duration $t_{FWHM}$, but as well on the square-root of the intensity[10]. Keeping in mind that the laser pulse is steepened in the CNF-plasma, this scaling would suggest a factor of slightly lower than 5.8 for the intensity increase. The optimum case of 8-fold intensity increase observed in the 3D simulation has not been reached in our experiment, but the trend with increasing length of propagation through the CNT-plasma is evident.

# Figures

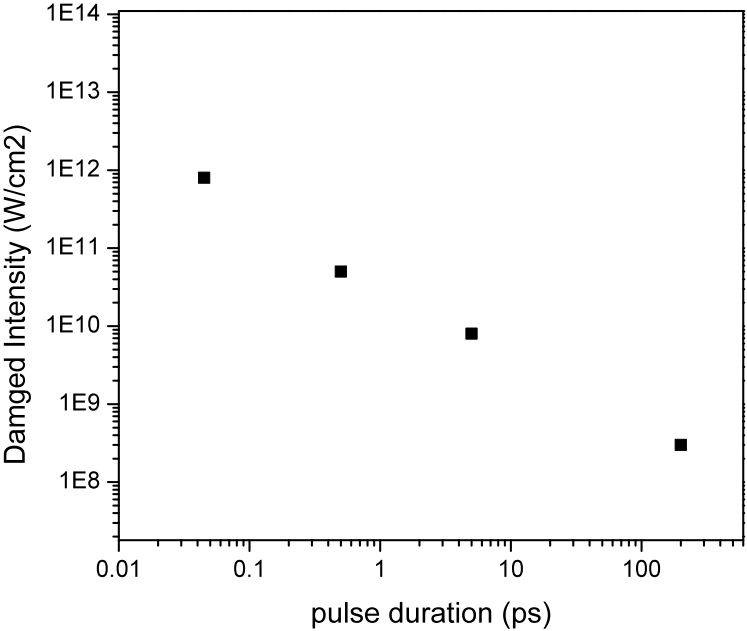

**Supplementary Figure 1**. Measurements of damage threshold for CNF with varying laser pulse duration from 30 fs to 200 ps.

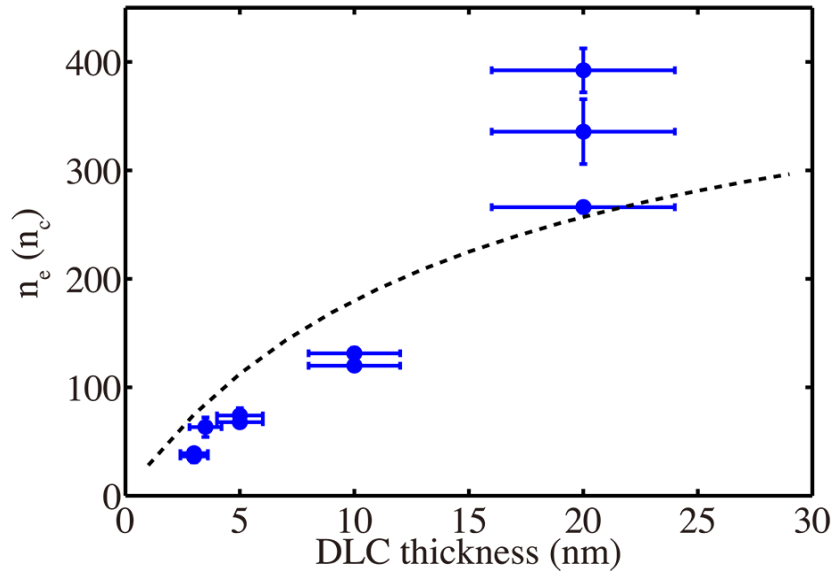

**Supplementary Figure 2**. Estimated effective density $n_e$ for different thicknesses of DLC foils derived from the laser transmission measurements.